# Electrical Control of Coherent Spin Rotation of a Single-Spin Qubit

Xiaoche Wang[1], Yuxuan Xiao[2], Chuanpu Liu[3], Eric Lee-Wong[1,4], Nathan J. McLaughlin[1], Hanfeng Wang[1], Mingzhong Wu[3], Hailong Wang[2], Eric E. Fullerton[2,4], Chunhui Rita Du[1,2]

[1]Department of Physics, University of California San Diego, La Jolla, California 92093
[2]Center for Memory and Recording Research, University of California, San Diego, La Jolla, California 92093
[3]Department of Physics, Colorado State University, Fort Collins, Colorado 80523
[4]Department of NanoEngineering, University of California, San Diego, La Jolla, California 92093

Nitrogen vacancy (NV) centers, optically-active atomic defects in diamond, have attracted tremendous interest for quantum sensing, network, and computing applications due to their excellent quantum coherence and remarkable versatility in a real, ambient environment. One of the critical challenges to develop NV-based quantum operation platforms results from the difficulty to locally address the quantum spin states of individual NV spins in a scalable, energy-efficient manner. Here, we report electrical control of the coherent spin rotation rate of a single-spin qubit in NV-magnet based hybrid quantum systems. By utilizing electrically generated spin currents, we are able to achieve efficient tuning of magnetic damping and the amplitude of the dipole fields generated by a micrometer-sized resonant magnet, enabling electrical control of the Rabi oscillation frequency of NV spins. Our results highlight the potential of NV centers in designing functional hybrid solid-state systems for next-generation quantum-information technologies. The demonstrated coupling between the NV centers and the propagating spin waves harbored by a magnetic insulator further points to the possibility to establish macroscale entanglement between distant spin qubits.

The past decade witnessed significant progress in new approaches for information processing, such as quantum,[1] neuromorphic,[2,3] and non-von Neumann computing[4] whose research interest is fueled by the saturation of downscaling and speed of the conventional CMOS technology and, more generally, the energy use of information technologies. Among these potential candidates, quantum computing employs algorithms that rely on inherent quantum properties of microscopic matters such as coherence, superposition, and entanglement and serves as a transformative operation platform enabling massively parallel processing of information in a compact physical system.[5] Many promising quantum systems including superconducting Josephson junctions,[6] topological insulators,[7] and trapped ions[8] have been extensively explored toward this end.

Nitrogen-vacancy (NV) centers,[9] optically-active atomic defects in diamond that act as single-spin quantum bits, are naturally relevant in this context. Due to their excellent quantum coherence time,[9] local spin-entanglement,[10] and notable versatility in a wide temperature range,[11,12] NV centers offer a remarkable platform to design emerging quantum architectures.[13–15] They have been successfully applied to quantum sensing, imaging, and quantum networks, exhibiting unprecedented field sensitivity,[9] spatial resolution,[16] and long-range photon-mediated qubit transmission.[17] Despite the remarkable progress, the role of NV centers in quantum computing has been rather peripheral. One of the major bottlenecks results from difficulties to locally address individual NV spin states in a scalable, energy-efficient manner. Presently, the quantum spin states of an individual NV center is mainly controlled by microwave fields generated by electric currents in a proximal metallic stripline or waveguide.[18] The dispersive Oersted fields slowly decay in real space, which imposes an inherent challenge to achieve excellent scalability in NV-based quantum operation systems. In addition, this approach typically requires a high microwave current density and the associated Joule heat can lead to decoherence of the quantum spin states.[19] While alternative approaches such as mechanical resonators,[20–22] magnetoelastic interaction,[23] and strain[20–22] have been actively explored recently, the desirable scalability or coupling strength remain to be lacking in these schemes.

In this article, we report energy-efficient, electrical control of single-spin rotation in a hybrid NV-magnet quantum system. The resonant spin waves excited in a proximal magnetic insulator, yttrium iron garnet (YIG), effectively amplify the amplitude of local microwave fields at the NV site, giving rise to orders of magnitude enhancement of the NV spin rotation rate. By utilizing the spin-orbit torque (SOT) generated by an adjacent platinum (Pt) layer,[24–26] we further demonstrated an efficient tuning of the magnetic damping of the resonant spin waves, enabling electrical control of spin rotation of a single-spin quantum qubit. We note that the mutual interactions between spin currents, magnetic devices, and NV spin qubits could be controlled in a scalable fashion down to a nanoscale regime, offering a new route to develop NV-based hybrid quantum computing platforms.

We first discuss the measurement system and device structure as illustrated in Fig. 1(a). A 10-μm-wide and 50-μm-long YIG/Pt strip was fabricated by standard photolithography and ion mill etching processes from a $Gd_3Ga_5O_{12}$ (substrate)/YIG (20 nm)/Pt (10 nm) film. A diamond nanobeam[27] containing individually addressable NV centers was transferred onto the surface of the YIG/Pt strip to establish nanoscale proximity between NV spins and the studied samples (see supplementary note 1). A 500-nm-thick on-chip Au stripline was fabricated next to the YIG/Pt strip, delivering microwave currents to manipulate the NV spin state and to excite ferromagnetic resonance (FMR) of the YIG strip. We employed a scanning confocal microscope to optically locate NV centers. A photoluminescence (PL) image shown in Fig. 1(b) provides an overview of the device structure, where an NV center ($NV_1$) is positioned on top of the YIG/Pt strip, demonstrating the single-spin addressability. We first performed optically detected magnetic resonance (ODMR) measurements to examine the NV electron spin resonance (ESR) and the FMR of the patterned YIG strip. A green laser was applied to constantly initiate the NV spin to the $m_s=0$ state, and the emitted PL was monitored via a single-photon detector. An external magnetic field $B_{ext}$ was applied along the NV-axis, with an angle of 60 degree relative to the $z$ axis as illustrated in Fig. 1(a). Figure 1(c) shows the normalized PL intensity as a function of the microwave frequency $f$ and the external magnetic field $B_{ext}$. The straight line denoted by $f_-$ results from the expected decrease in NV fluorescence of NV ESR transition in the electronic ground state: $f_- = 2.87 - \gamma B_{ext}/2\pi$, where $\gamma$ is the gyromagnetic ratio. The other two straight lines denoted by $f_{e\pm}$ result from the NV ESR at the optically excited state: $f_{e\pm} = 1.42 \pm \gamma B_{ext}/2\pi$. The NV fluorescence also decreases when $f$ matches the FMR frequency $f_{FMR}$ of the YIG strip as shown by the curved dash line below $f_{e+}$. This NV-based off-resonant detection of spin wave modes in a proximal ferromagnet is attributed to the multi-magnon scattering processes, giving rise to enhanced magnon densities at the NV ESR frequencies.[28,29]

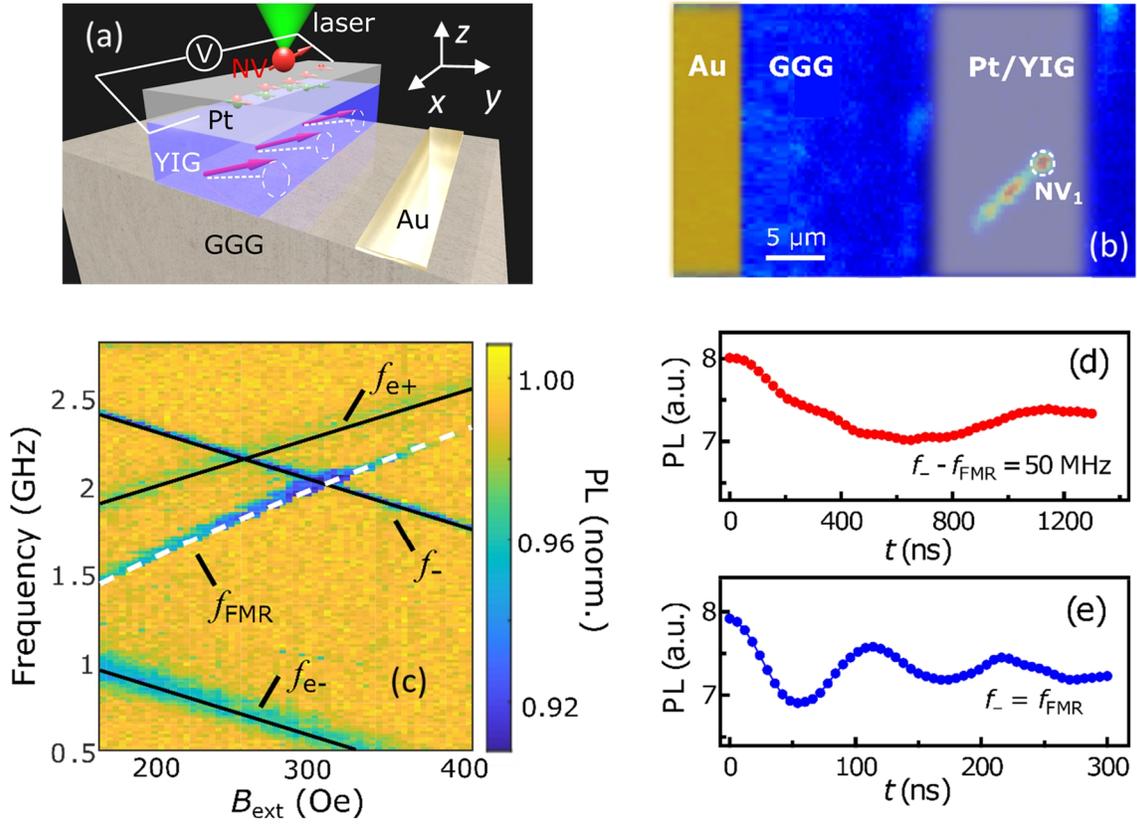

**Figure 1:** (a) Schematic of an NV spin positioned on top of a patterned YIG (20 nm)/Pt (10 nm) strip to investigate the amplification and electrical tuning effect of the NV spin rotation rate at the YIG FMR condition. (b) A photoluminescence image showing an overview of a fabricated device consisting of an Au stripline (false yellow color) and a diamond nanobeam containing individually addressable NV spins on top of the YIG/Pt strip (false-silver color). (c) Normalized PL intensity of $NV_1$ measured as a function of $B_{ext}$ and microwave frequency $f$. The black straight lines represent the NV ESR transitions in the electronic ground (excited) state and the white dash line represents the field dispersion curve of the YIG FMR mode. The time dependent PL intensity of $NV_1$ measured at (d) $f_- = f_{FMR} + 50$ MHz and (e) $f_- = f_{FMR}$.

Next, we performed NV Rabi oscillation measurements to characterize the coherent spin rotation rate $f_{Rabi}$ of NV$_1$. When a microwave magnetic field with NV ESR frequencies $f_\pm$ is applied at the NV site, the NV spin will periodically oscillate between two different states, *i.e.* m$_s$ = 0 and $\pm 1$ in the rotation frame, which is referred to as Rabi oscillations.[30] Here, $f_\pm$ characterize NV ESR frequencies corresponding to the spin transition between m$_s$ = 0 and m$_s$ = $\pm 1$ states. The coherent spin rotation rate $f_{Rabi}$ is proportional to the amplitude of the local microwave field that is perpendicular to the NV-axis.[31] Figures 1(d) and 1(e) show the measured PL intensity of NV$_1$ as a function of the microwave duration time $t$ at two different NV ESR frequencies. When $f_-$ is detuned from $f_{FMR}$ by 50 MHz, the measured PL spectrum slowly oscillates with a characteristic $f_{Rabi}$ of 0.8 MHz, from which the local microwave field $h_{rf}$ generated by the Au stripline is estimated to be 0.5 Oe. When $f_- = f_{FMR}$, notably, the NV PL spectrum exhibits a significantly accelerated oscillation behavior with an enhancement of $f_{Rabi}$ from 0.8 to 9 MHz. This one order of magnitude enhancement of the NV coherent spin rotation rate results from a larger oscillating stray field $h_{FMR}$ generated by the quasi-uniform precession of the YIG magnetization, which amplifies the effective microwave magnetic field experienced by the NV spin (see supplementary note 2).

To achieve electrical control of the coherent NV spin rotation rate, we further employed the SOT generated by the Pt layer to vary the amplitude of the oscillating stray field $h_{FMR}$ generated by the resonant YIG strip. Figure 2(a) illustrates the optical, microwave, and electrical measurement sequence. A 3-μs-long green laser pulse was first applied to initialize the NV spin to the m$_s$ = 0 state. A microwave pulse at a frequency $f_-$ was applied to induce an m$_s$ = 0 ↔ -1 transition. To minimize the current-induced Joule heat, an electric current sequence synchronized with the microwave pulse was applied in the Pt layer. Last, a second green laser pulse was applied to measure the spin-dependent PL of the NV center and re-initialize the NV spin for the next measurement sequence. The time duration of the microwave (electrical) pulses systematically varies from zero to a few hundred nanoseconds in order to detect a time-dependent variation of the NV PL intensity. Figure 2(b) shows the Rabi oscillation spectrum of NV$_1$ measured at three different electric current densities $J_c$ when $f_- = f_{FMR}$. Without applying an electric current, $f_{Rabi}$ is measured to be 9 MHz, exhibiting a significant enhancement in comparison to the off-FMR condition as discussed above. When $J_c = -1 \times 10^{11}$ A/m$^2$, we observed a faster oscillation behavior of the measured PL spectrum with an enhanced $f_{Rabi}$ = 11 MHz. When $J_c = 1 \times 10^{11}$ A/m$^2$, the NV spin exhibits a slower oscillation behavior with a reduced $f_{Rabi}$ of 7 MHz. To further illustrate the electrical tuning and amplification effect at the YIG FMR condition, Fig. 2(c) plots the normalized NV Rabi frequency ($f_{Rabi}/\sqrt{P}$) as a function of $f_- - f_{FMR}$ when $J_c = 0$ and $\pm 1 \times 10^{11}$ A/m$^2$. Note that the variation of the input microwave power $P$ needs to be normalized to characterize the driving efficiency of NV spin rotation (see supplementary note 3). Figure 2(d) plots $f_{Rabi}/\sqrt{P}$ measured at $f_- = f_{FMR}$ as a function of the electric current density. In general, $f_{Rabi}/\sqrt{P}$ follows a quasi-linear dependence on $J_c$ and exhibits ~$\pm 23\%$ variation when $J_c = \pm 1 \times 10^{11}$ A/m$^2$.

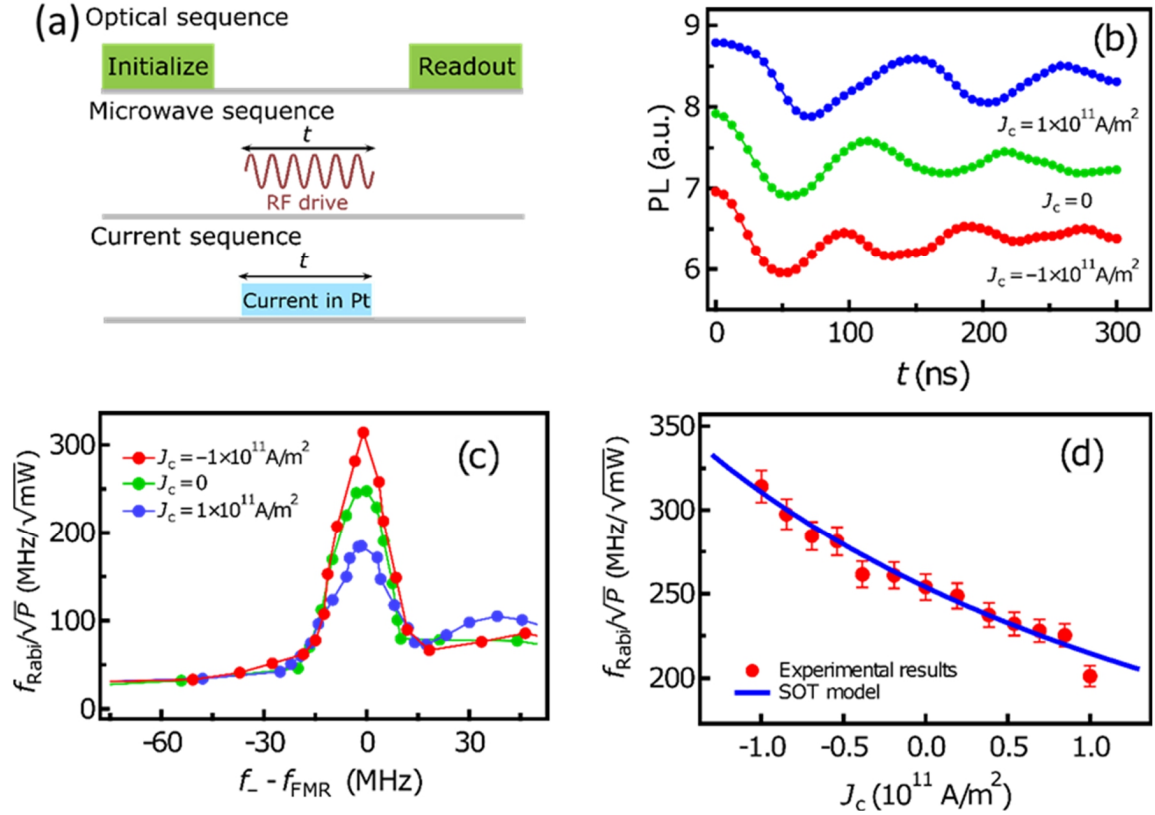

**Figure 2:** (a) Schematic of the measurement sequence to electrically control NV Rabi frequency. (b) The time dependent PL intensity of $NV_1$ measured at $J_c = 0$ and $\pm 1 \times 10^{11}$ A/m². (c) Normalized NV Rabi frequency $f_{\text{Rabi}}/\sqrt{P}$ measured as a function of $f_- - f_{\text{FMR}}$. The green, blue, and red curves correspond to $J_c = 0$ and $\pm 1 \times 10^{11}$ A/m², respectively. (d) When $f_- = f_{\text{FMR}}$, the measured current density dependence of $f_{\text{Rabi}}/\sqrt{P}$ (red points) agrees well with the theoretical prediction based on the SOT-model (blue line).

The electrically tunable $f_{\text{Rabi}}$ results from the SOT-induced variation of the local microwave magnetic field at the NV site. When an electric current flows through the Pt layer, a spin current $J_s$ will be generated by the spin Hall effect.[32] Due to the interfacial scattering process, $J_s$ will transport across the YIG/Pt interface and thereby exerts a damping-like spin-transfer torque $\boldsymbol{\tau} = \boldsymbol{m} \times (\boldsymbol{m} \times \boldsymbol{s})$ on the YIG magnetization. Here, $\boldsymbol{m}$ is the magnetization of the YIG pattern and $\boldsymbol{s}$ is the spin polarization of the injected spin currents. Depending on the polarity of the electric current, the magnitude of $\boldsymbol{\tau}$ effectively increases or decreases the precessional cone angle $\Theta$ of the YIG magnetization, leading to a variation of the amplitude of the oscillating stray field $h_{\text{FMR}}$ as follows: $h_{\text{FMR}} \propto M_s \sin\Theta$. According to the SOT model, the electric current density dependence of $\Theta$ is given by:[24,25]

$$f_{\text{Rabi}}/\sqrt{P} \propto \Theta = \frac{h_{\text{rf}}}{\Delta H_0 + \frac{2\pi f_{\text{FMR}}}{\gamma}\left(\alpha + \frac{\sin\varphi}{(B_{\text{ext}} + 2\pi M_{\text{eff}})\mu_0 M_s t_{\text{YIG}}} \frac{\kappa \theta_{\text{SH}} \hbar J_c}{2e}\right)} \quad (1)$$

where $\Delta H_0$ is the film inhomogeneity contribution to the FMR linewidth, $\mu_0$ is the free-space permeability, $\hbar$ is the reduced Planck constant, $\alpha$, $M_{\text{eff}}$, $M_s$, and $t_{\text{YIG}}$ correspond to the intrinsic magnetic damping, effective demagnetizing field, saturation magnetization, and thickness of the YIG strip, respectively. $\varphi$ characterizes the angle between the in-plane projection of the YIG magnetization and the applied electric current, $\kappa$ characterizes the spin transport efficiency at the YIG/Pt interface,[25] and $\theta_{SH}$ is the spin Hall angle of the Pt layer. Taking $\theta_{SH} = 0.07$,[32] $\kappa = 0.25$ with other known material parameters ($\Delta H_0 = 6.3$ Oe, $\alpha = 0.001$, and $M_s = M_{eff} = 1.31 \times 10^5$ A/m), the blue curve in Fig. 2(d) plots the current density dependence of $f_{Rabi}/\sqrt{P}$ predicted by the SOT-model, which is in excellent agreement with our experimental results.

So far, we have demonstrated the electrical control of coherent spin rotation of an NV single spin by the quasi-FMR spin wave mode of a proximal ferromagnet. Next, we further extend the measurement platform to a more general scenario, where propagating spin waves with certain wavevectors and group velocities are involved. Figure 3(a) shows the schematic of the device structure, where an Au coplanar waveguide (CPW) and an insulating SiO$_x$ spacer are fabricated on a patterned 80-μm-wide and 300-μm-long YIG (100 nm)/Pt (10 nm) waveguide. The width of the signal and ground lines of the CPW and the center-to-center separation between them are designed to be 1.5 μm and 3.15 μm, respectively, and the long-axis of the CPW is perpendicular to the YIG waveguide as shown by the scanning electron microscope image of Fig. 3(b). A diamond nanobeam containing an individual NV center (NV2) was transferred on top of the Pt layer with a distance of ~5.5 μm to the center of signal line. An external magnetic field $B_{ext}$ was applied along the long-axis of the CPW to excite the Damon-Eschbach surface spin wave mode[33] at the YIG/Pt interface. The propagating nature of the excited spin wave has been experimentally confirmed by the measurements of microwave transmission between the two CPWs (see supplementary note 4). Figure 3(c) plots the characteristic wavevector distribution spectrum. The well-defined excitation peaks are determined by the spatial distribution of the microwave fields generated by the CPW (see supplementary note 5).[34] Figure 3(d) shows the ODMR map measured by the NV spin sensor, where up to four propagating surface spin wave modes with distinct wavevectors: $k_1$, $k_2$, $k_3$, and $k_4$ emerge. The field dispersion curves of these spin wave modes follow the theoretical prediction as shown by the white dash lines (see supplementary note 5) and cross with the NV ESR frequency $f_-$ between 2.1 and 2.6 GHz.

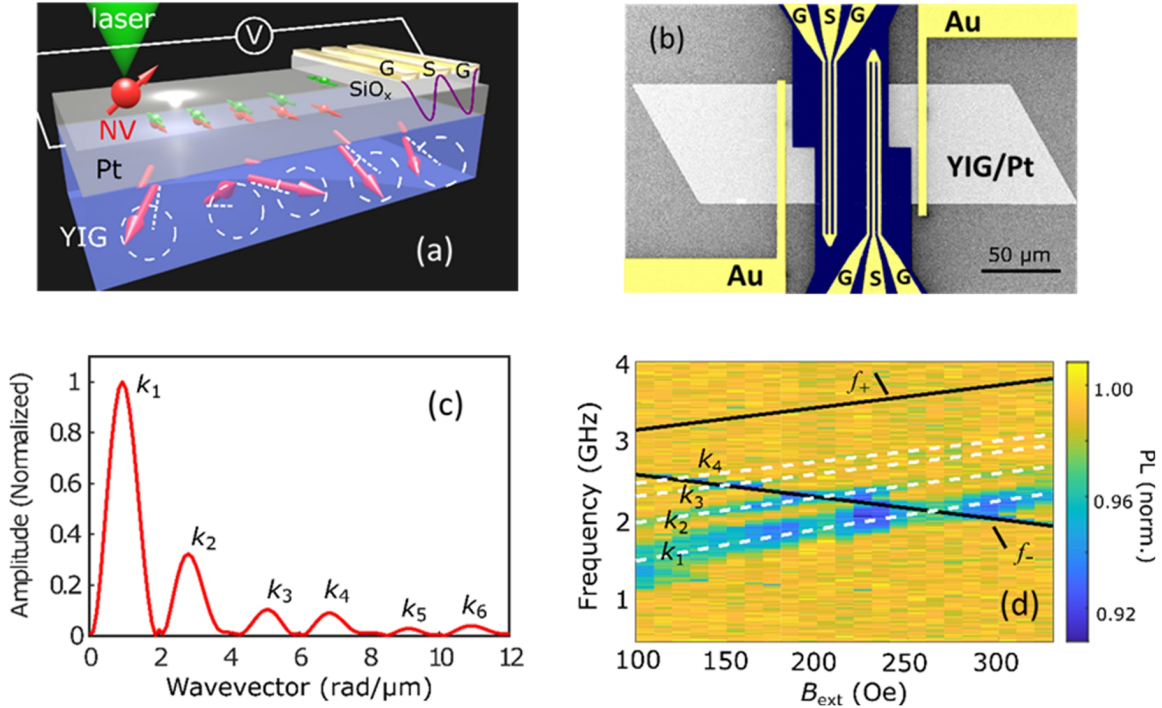

**Figure 3:** (a) Schematic of the NV-YIG/Pt device to establish a coherent coupling between propagating spin waves and the NV spin. (b) Scanning electron microscope image of a prepared device with a patterned YIG waveguide and two CPWs to excite propagating surface spin waves. An electric current flowing through the Pt layer is employed to generated spin currents, enabling control of magnetic damping and amplitude of the excited spin waves. G and S represent ground and signal lines, respectively. (c) Spin-wave excitation spectra obtained by Fourier transformation of the microwave fields generated by the CPW. (d) Normalized PL intensity of the NV center as a function of $B_{ext}$ and microwave frequency $f$. The two black straight lines correspond to the NV ESR transitions and the four white dash lines correspond to the field dispersion curves of propagating surface spin wave modes with characteristic wavevectors of $k_1$, $k_2$, $k_3$, and $k_4$, respectively.

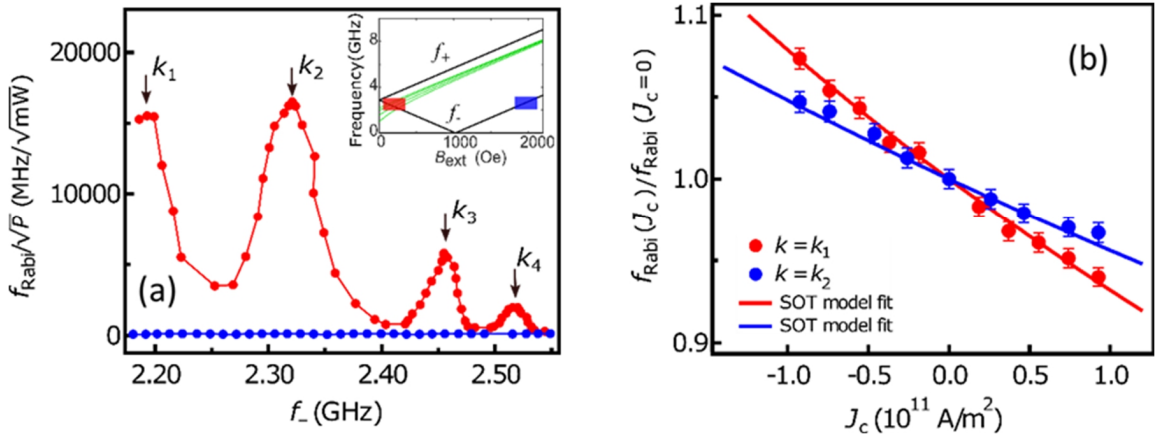

**Figure 4:** (a) Normalized NV spin rotation rate $f_{Rabi}/\sqrt{P}$ measured as a function of $f_-$ in the low magnetic field (red curve) and high magnetic field (blue curve) regimes. Inset: the variation of the NV ESR frequencies $f_\pm$ (black curves) and the resonant frequencies of propagating surface spin waves (green curves) as a function of $B_{ext}$. (b) $f_{Rabi}(J_c)/f_{Rabi}(J_c = 0)$ measured as a function of $J_c$ at the resonant condition of $k_1$ (red points) and $k_2$ (blue points) spin wave modes. The red and blue curves plot the theoretical prediction based on the SOT-model.

To illustrate the amplification effect of the propagating spin waves on the NV spin rotation rate, the red curve in Fig. 4(a) shows $f_{Rabi}/\sqrt{P}$ measured as a function of $f_-$. Remarkably, $f_{Rabi}/\sqrt{P}$ reaches the peak values of 15531, 16509, 5774, and 1977 MHz/$\sqrt{mW}$ when $f_-$ meets the resonant condition of the $k_1$, $k_2$, $k_3$, and $k_4$ spin wave modes, respectively. As a comparison, the blue curve shows $f_{Rabi}/\sqrt{P}$ measured in the high magnetic field regime where $f_-$ stays far away from the resonant frequencies of the series of spin wave modes as illustrated in the inset of Fig. 4(a). At the off-resonant condition, $f_{Rabi}/\sqrt{P}$ follows an average value of 120 MHz/$\sqrt{mW}$. The enhancement ratio of $f_{Rabi}/\sqrt{P}$ reaches 129, 138, 48, and 17 at the resonant condition of $k_1$, $k_2$, $k_3$, and $k_4$ spin wave modes, respectively, in quantitative agreement with the analytical theoretical calculations[35] (see supplementary note 2). We notice that the enhancement of $f_{Rabi}/\sqrt{P}$ decreases with the increase of the wavevector, which is attributable to a reduced microwave excitation efficiency of the higher order propagating spin wave modes.[36] Similar to the quasi-FMR spin wave mode, we also employ spin currents generated by the Pt layer as a tuning knob to electrically control the NV spin rotation rate. Figure 4(b) plots the variation of $f_{\text{Rabi}}/\sqrt{P}$ as a function of $J_c$ at the resonant condition of the $k_1$ and $k_2$ spin wave modes. The normalized NV Rabi frequencies exhibit a systematic variation on the magnitude of $J_c$, in agreement with the SOT model predicted by Eq. (1). The reduced electrical tunability of NV spin rotation rate results from a much larger thickness of the YIG film as well as the higher resonant frequencies and wave vector of the propagating spin wave modes.[37] In addition to the surface spin wave modes, we also observed the similar behaviors for back volume spin wave modes when the external magnetic field is parallel to the short-axis of the CPWs, confirming the universality of the coupling between NV spins and the propagating magnons (see supplementary note 6).

In summary, we have demonstrated electrical control of the coherent spin rotation rate of a single-spin qubit in an NV-magnet quantum system. By applying an electric current in a YIG/Pt nanostructure, we observed a spin-current-induced variation of the Rabi frequency when the NV ESR frequency meets the resonant condition of spin wave modes. Further shrinking the dimension of the magnetic devices to sub-micrometer regime, where the generated spin currents could completely compensate the damping of the ferromagnet and excite the auto-oscillations,[38] the spin state of NV centers could be fully electrically addressed in absence of external microwave currents. We note that excellent quantum coherence is still preserved in NV centers in this process. The measured spin coherent time is one order of magnitude larger than the reported value of nanodiamonds[14] and comparable to NV spins contained in bulk diamond structures (see supplementary note 7),[39] which promises for applications such as sensitivity metrology,[40] quantum computing,[13] and communications.[17] The demonstrated dipole coupling between single the NV spin and the propagating spin waves also serves as an ideal medium to establish long-range entanglement between distant NV spin qubits,[41] offering a new opportunity in designing an NV-based quantum operation platform.

## Methods

**Materials and device fabrication.** The 20-nm thick $Y_3Fe_5O_{12}$ (YIG) films used in this work were deposited by magnetron sputtering on (111)-oriented $Gd_3Ga_5O_{12}$ (GGG) substrates. The saturation magnetization is measured to be $1.31 \times 10^5$ A/m. The details of the growth parameters have been reported in the previous work.[42] The 100-nm thick YIG films were grown by liquid-phase epitaxy method and they were commercially available from the company Matesy GmbH. The saturation magnetization is measured to be $1.35 \times 10^5$ A/m. Two types of nitrogen-vacancy (NV)-YIG/Pt devices prepared by standard photolithography, ion beam etching, and sputtering processes. For the device illustrated in Fig. 1(a), a 10-μm-wide and 50-μm-long YIG (20 nm)/Pt (10 nm) strip was first defined on a GGG substrate, followed by the fabrication of a 19.6-μm-wide and 500-nm thick Au stripline. For the device illustrated in Fig. 3(a), a 80-μm-wide and 300-μm-long YIG (100 nm)/Pt (10 nm) waveguide was first created and Au coplanar waveguides (CPWs) were fabricated on the defined YIG waveguide with the perpendicular orientation. Patterned diamond nanobeams containing NV centers were picked up and transferred onto the magnetic nanostructures using a tungsten tip performed under a micromechanical transfer stage. Nanobeams were fabricated by a combination of top-down etching and angled etching processes.[27] Acid cleaning was performed before and after the fabrication processes to ensure the pristinity the diamond surface, which is crucial to establish nanoscale proximity between NVs and the studied samples.

**NV measurements.** NV measurements were performed by a home-built scanning confocal microscope. Green laser pulses used for the NV initiation and readout were generated by an acoustic optical modulator with a double-pass configuration. NV spin state was optically addressed through integrating the measured photoluminescence (PL) generated during the first 600 ns of the green laser readout pulse. NV Rabi oscillation measurements were performed using the sequence shown in Fig. 2(a) in the main text. The microwave signals were generated by a Rohde & Schwarz SGS100a and an Agilent N9310A and connected to a microwave switch (Minicircuits ZASWA-2-50DR+). The pulses to trigger the modulation of microwave amplitude (on and off) were generated by an arbitrary waveform generator (Tektronix AWG5014C). It was also used to apply the synchronized electrical pulses in the Pt layer to minimize the thermal heating effect (see supplementary note 8). The trigger pulses to the optical modulator and photon counting were generated by a programmable pulse generator (Spincore, PBESR-PRO-500).

**COMSOL Multiphysics simulations.** "Electromagnetic Waves, Frequency Domain" module was used to simulate the spatial profile of the microwave magnetic fields generated by an Au CPW on top of a YIG thin film. Electric currents following through the signal and ground lines of the CPW were set to be $+I$, $-I/2$, and $-I/2$, respectively. Equilateral triangular meshes with varying dimensions were used in the simulations. In the area that is in vicinity of the CPW, fine meshes with a length of 0.4 μm was used. We set a growth rate of the mesh size to have large meshes with a length of 40 μm in the area that is far away from the CPW. By solving the Helmholtz Equation, we could obtain the distribution of the magnetic fields in real space. The spin-wave excitation spectra in the momentum space was extracted via Fourier transformation of the magnetic field profile (see supplementary note 5).

**Data availability**. All data supporting the findings of this study are available from the corresponding author on reasonable request.

**Acknowledgements**:
We thank Feiyang Ye for helpful discussions to improve the manuscript. Y. X., H. W., and E. E. F. were supported by Quantum-Materials for Energy Efficient Neuromorphic-Computing, an Energy Frontier Research Center funded by DOE, Office of Science, BES under Award NO. DE-SC0019273. C. L. and M. W. were supported by the U.S. National Science Foundation (EFMA-1641989 and ECCS-1915849). This work was also supported by the Startup Funds from UCSD.

**Author contributions**
C. D. conceived the idea and designed the project. X. W. performed the measurements. E. L.-W. and N. M. built the confocal setup and performed the diamond nanobeam transfer. X. Y., H. L. W., and E. E. F. fabricated the devices. C. L. and M. W. provided the YIG samples. H. F. W. contributed to the COMSOL Multiphysics simulations. H. L. W. and C. D. wrote the manuscript with the help from all co-authors.

**Competing interests**
The authors declare no competing interests.

**Reference:**
1. Nielsen, M. A., & Chuang, I. L. *Quantum Computation and Quantum Information* (Cambridge Univ. Press, Cambridge, 2000).
2. Monroe, D. Neuromorphic computing gets ready for the (really) big time. *Commun. ACM* **57**, 13 (2014).
3. del Valle, J. *et al.* Subthreshold firing in Mott nanodevices. *Nature* **569**, 388 (2019).
4. Von Neumann, J. First draft of a report on the EDVAC, *IEEE Ann. Hist. Comput.* **15**, 27 (1993).
5. Bravyi, S., Gosset, D., & König, R. Quantum advantage with shallow circuits. *Science,* **362**, 308 (2018).
6. Devoret, M. H., & Schoelkopf, R. J. Superconducting circuits for quantum information: An Outlook. *Science* **39**, 1169 (2013).
7. Kitaev, A. Y. Fault-tolerant quantum computation by anyons. *Ann. Phys.* **303**, 2 (2003).
8. Cirac, J. I., & Zoller, P. Quantum computations with cold trapped ions. *Phys. Rev. Lett.* 74, 4091 (1995).
9. Rondin, L., Tetienne, J.-P., Hingant, T., Roch, J.-F., Maletinsky, P., & Jacques, V. Magnetometry with nitrogen-vacancy defects in diamond. *Rep. Prog. Phys.* **77**, 56503 (2014).
10. Dolde, F., Jakobi, I., Naydenov, B., Zhao, N., Pezzagna, S., Trautmann, C., Meijer, J., Neumann, P., Jelezko, F., & Wrachtrup, J. Room-temperature entanglement between single defect spins in diamond. *Nat. Phys.* **9**, 139 (2013).
11. Liu, G. Q., Feng, X., Wang, N., Li, Q., & Liu, R. B. Coherent quantum control of nitrogen-vacancy center spins near 1000 kelvin. *Nat. Commun.* **10**, 1344 (2019).
12. Pelliccione, M., Jenkins, A., Ovartchaiyapong, P., Reetz, C., Emmanouilidou, E., Ni, N., & Jayich, A. C. B. Scanned probe imaging of nanoscale magnetism at cryogenic temperatures with a single-spin quantum sensor. *Nat. Nanotechnol.* **11**, 700 (2016).
13. Childress, L., & Hanson, R. Diamond NV centers for quantum computing and quantum networks. *MRS Bulletin* **38**, 134 (2013).
14. Andrich, P. *et al.* Long-range spin wave mediated control of defect qubits in nanodiamonds. *npj Quantum Inf.* **3**, 28 (2017).
15. Kikuchi, D. *et al.* Long-distance excitation of nitrogen-vacancy centers in diamond via surface spin waves. *Appl. Phys. Express* **10**, 103004 (2017).
16. Grinolds, M. S. *et al.* Subnanometre resolution in three-dimensional magnetic resonance imaging of individual dark spins. *Nat. Nanotechnol.* **9**, 279 (2014).
17. Hensen, B. *et al*. Loophole-free Bell inequality violation using electron spins separated by 1.3 kilometres. *Nature* **526**, 682 (2015).
18. Fuchs, G. D., Dobrovitski, V. V., Toyli, D. M., Heremans, F. J., & Awschalom, D. D. Gigahertz dynamics of a strongly driven single quantum spin. *Science* **326**, 1520 (2009).
19. Jarmola, A., Acosta, V. M., Jensen, K., Chemerisov, S., & Budker, D. Temperature- and magnetic-field-dependent longitudinal spin relaxation in nitrogen-vacancy ensembles in diamond. *Phys. Rev. Lett.* **108**, 197601 (2012).
20. Barfuss, A., Teissier, J., Neu, E., Nunnenkamp, A., & Maletinsky, P. Strong mechanical driving of a single electron spin. *Nat. Phys.* **11**, 820 (2015).
21. MacQuarrie, E. R., Gosavi, T. A., Jungwirth, N. R., Bhave, S. A., & Fuchs, G. D., Mechanical spin control of nitrogen-vacancy centers in diamond. *Phys. Rev. Lett.* **111**, 227602 (2013).
22. Ovartchaiyapong, P., Lee, K. W., Myers, B. A., & Jayich, A. C. B. Dynamic strain-mediated coupling of a single diamond spin to a mechanical resonator. *Nat. Commun.* **5**, 4429 (2014).

23. Labanowski, D. *et al.* Voltage-driven, local, and efficient excitation of nitrogen-vacancy centers in diamond. *Sci. Adv.* **4**, eaat6574 (2018).
24. Liu, L., Moriyama, T., Ralph, D. C., & Buhrman, R. A., Spin-torque ferromagnetic resonance induced by the spin Hall effect. *Phys. Rev. Lett.* **106**, 036601 (2011).
25. Hamadeh, A. *et al.* Full control of the spin-wave damping in a magnetic insulator using spin-orbit torque. *Phys. Rev. Lett.* **113**, 197203 (2014).
26. Solyom, A. *et al.* Probing a Spin Transfer Controlled Magnetic Nanowire with a Single Nitrogen-Vacancy Spin in Bulk Diamond. *Nano Lett.* **18**, 6494 (2018).
27. Burek, M. J., de Leon, N. P., Shields, B. J., Hausmann, B. J. M., Chu, Y., Quan, Q., Zibrov, A. S., Park, H., Lukin, M. D., & M. Lončar, M. Free-standing mechanical and photonic nanostructures in single-crystal diamond. *Nano Lett.* **12**, 6084 (2012).
28. Du, C. H. *et al.* Control and Local Measurement of the Spin Chemical Potential in a Magnetic Insulator. *Science* **357**, 195 (2017).
29. Wolfe, C. S., Bhallamudi, V. P., Wang, H. L., Du, C. H., Manuilov, S., Teeling-Smith, R. M., Berger, A. J., Adur, R., Yang, F. Y., & Hammel, P. C. Off-resonant manipulation of spins in diamond via precessing magnetization of a proximal ferromagnet. *Phys. Rev. B* **89**, 180406 (2014).
30. Jelezko, F., Gaebel, T., Popa, I., Gruber, A., & Wrachtrup, J. Observation of coherent oscillations in a single electron spin. *Phys. Rev. Lett.* **92**, 076401 (2004).
31. Taminiau, T. H., Cramer, J., van der Sar, T., Dobrovitski, V. V., & Hanson, R. Universal control and error correction in multi-qubit spin registers in diamond. *Nat. Nanotechnol.* **9**, 171 (2014).
32. Liu, L., Pai, C.-F., Li, Y., Tseng, H. W., Ralph, D. C., & Buhrman, R. A. Spin-torque switching with the giant spin Hall effect of Tantalum. *Science* **336**, 555 (2012).
33. An, T. *et al.* Unidirectional spin-wave heat conveyer. *Nat. Mater.* **12**, 549 (2013).
34. Qin, H., Hämäläinen, S. J., Arjas, K., Witteveen, J., & van Dijken, S. Propagating spin waves in nanometer-thick yttrium iron garnet films: Dependence on wave vector, magnetic field strength, and angle. *Phys. Rev. B* **98**, 224422 (2018).
35. Mühlherr, C., Shkolnikov, V. O., & Burkard, G. Magnetic resonance in defect spins mediated by spin waves. *Phys. Rev. B* **99**, 195413 (2019).
36. Yu, H. *et al.* Magnetic thin-film insulator with ultra-low spin wave damping for coherent nanomagnonics. *Sci. Rep.* **4**, 6848 (2014).
37. Li, Y. & Bailey, W. E. Wave-number-dependent Gilbert damping in metallic ferromagnets. *Phys. Rev. Lett.* **116**, 117602 (2016).
38. Safranski, C. *et al.* Spin caloritronic nano-oscillator. *Nat. Commun.* **8**, 117 (2017).
39. De Lange, G., Wang, Z. H., Ristè, D., Dobrovitski, V. V., & Hanson, R. Universal dynamical decoupling of a single solid-state spin from a spin bath. *Science* **330**, 60 (2010).
40. Degen, C. L., Reinhard, F., & Cappellaro, P. Quantum sensing. *Rev. Mod. Phys.* **89**, 035002 (2017).
41. Trifunovic, L., Pedrocchi, F. L., & Loss, D. Long-distance entanglement of spin qubits via ferromagnet. *Phys. Rev. X* **3**, 1 (2014).
42. Sun, Y. *et al.* Damping in yttrium iron garnet nanoscale films capped by platinum. *Phys. Rev. Lett.* **111**, 106601 (2013).